\documentclass[12pt]{iopart}

\usepackage{iopams}  
\usepackage{epsfig}

\begin{document}

\title{Photons and Dileptons at LHC}

\author{Rainer J Fries$^1$, Simon Turbide$^2$, Charles Gale$^2$
and Dinesh K Srivastava$^3$}
\address{$^1$ Cyclotron Institute and Department of Physics,
Texas A\&M University, College Station TX 77843, USA}
\address{$^1$ RIKEN/BNL Research Center, Brookhaven National Laboratory, 
Upton NY 11973}
\address{$^2$ Department of Physics, McGill University, Montr\'eal, 
  Canada H3A 2T8}
\address{$^3$ Variable Energy Cyclotron Center, 1/AF Bidhan Nagar, 
Kolkata 700 064, India}

\begin{abstract}
We discuss real and virtual photon sources in heavy ion collisions and 
present results for dilepton yields in Pb+Pb collisions at the LHC
at intermediate and large transverse momentum $p_T$.
\end{abstract}


Electromagnetic radiation provides a valuable tool to understand the dynamics
of heavy ion collisions. Due to their long mean free path real and virtual 
photons carry information from very early times and from deep inside the 
fireball. We discuss the sources of photons which will be important for
the upcoming heavy ion experiments at LHC. We focus on intermediate and large 
transverse momenta $p_T$ and masses $M$. We also present our numerical results 
for dilepton yields.

At asymptotically large $p_T$ the most important source of real and virtual 
photons is the direct hard production in primary parton-parton collisions
between the nuclei, via Compton scattering, annihilation, and the 
Drell-Yan processes. These photons do not carry any signature of the
fireball. They are augmented by photons fragmenting from hard jets also
created in primary parton-parton collisions. The emission of this vacuum 
bremsstrahlung is described by real and virtual photon fragmentation 
functions. Vacuum fragmentation is assumed to happen outside the fireball,
so the jets are subject to the full energy loss in the medium. This 
contribution to the photon and dilepton yield is therefore depleted in
heavy ion collisions analogous to the high-$p_T$ hadron yield.

At intermediate scales jet-induced photons from the medium become 
important. It has been shown that high-$p_T$ jets interacting with 
the medium can produce real and virtual photons by one of two processes: 
(i) by Compton scattering or annihilation with a thermal parton, leading 
to an effective conversion of the jet into a photon \cite{Fries:2002kt}; 
(ii) by medium induced Bremsstrahlung \cite{Zakharov:2004bi}. 
Jet-medium photons have a steeper spectrum than primary photons and 
carry information about the temperature of the medium. They are also 
sensitive to the partial energy loss that a jet suffers from its creation 
to the point of emission of the photon.
At even lower $p_T$ and $M$ thermal radiation from the quark gluon plasma
(and also the hadronic phase not considered here) has to be taken into
account.

Figure 1 shows numerical evaluations of the different contributions discussed
above to the e$^+$e$^-$ transverse momentum and mass spectrum for central 
Pb+Pb collisions at LHC. We use next-to-leading order pQCD calculations for 
Drell Yan and a leading order calculation for jet production. Energy 
loss of jets is computed with the AMY formalism \cite{Turbide:2005fk}. 
Jet-medium emission and thermal emission have been evaluated in the Hard 
Thermal Loop (HTL) resummation scheme. For the mass spectrum we also show 
the expected background from correlated heavy quark decays. The full 
calculation for dileptons with a more extended discussion is presented in 
\cite{Turbide:2006mc}. Predictions for direct photon yields including 
jet-medium photons can be found in \cite{Turbide:2005fk}.

Dileptons from jet-medium interactions will be more important at LHC than
at previous lower energy experiments. They will be as important or even 
exceeding the Drell-Yan yields at intermediate masses up to about 8 GeV.
They offer a new way to access information about the temperature and the
partonic nature of the fireball.

\begin{figure}
\begin{center}
\epsfig{file=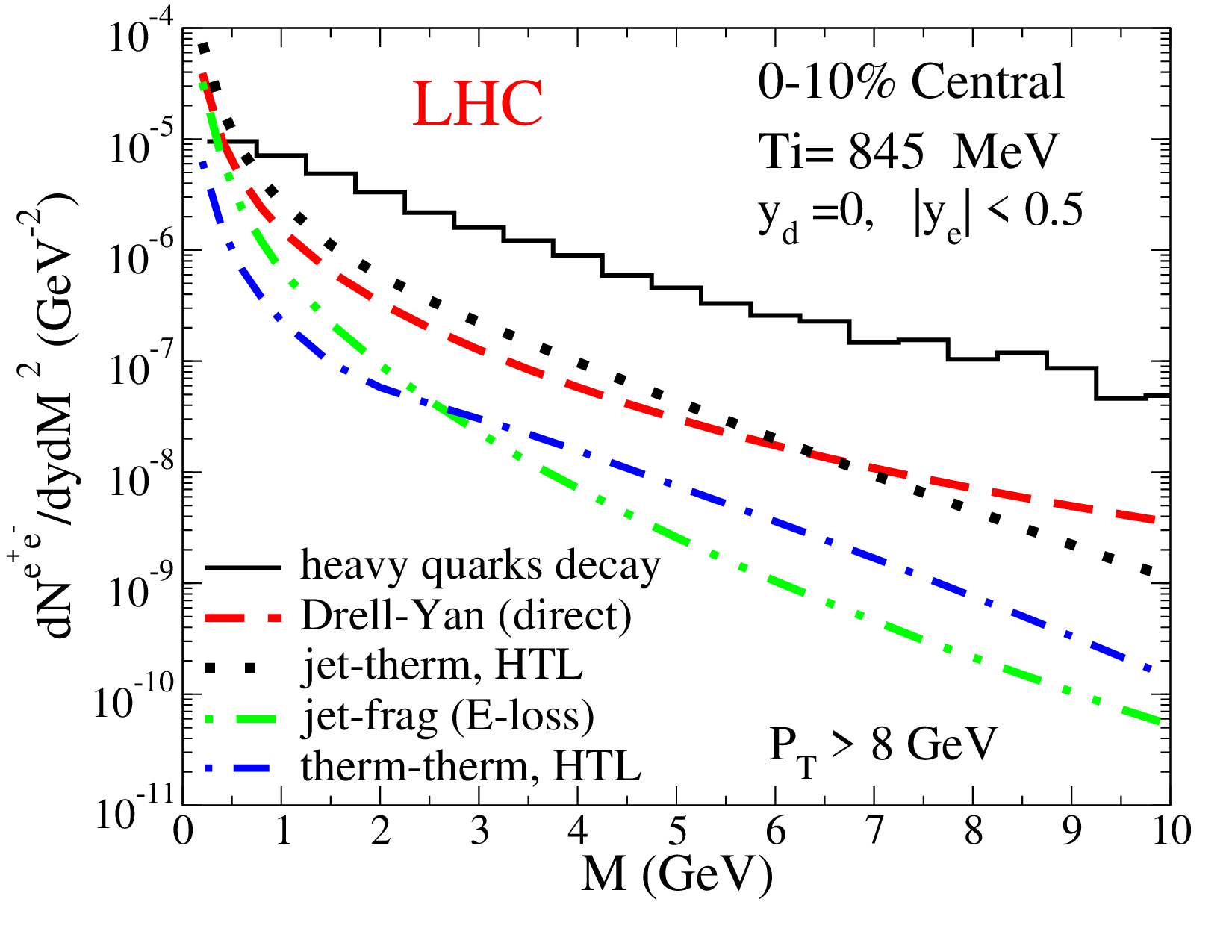,width=7.3cm} 
\epsfig{file=dilep_dpt_lowm_lhc_K_sum.eps,width=7.5cm}
\caption{The yield of e$^+$e$^-$ pairs in central Pb+Pb collisions at 
$\sqrt{s_{\mathrm{NN}}} =5.5$ TeV. 
{\it Left}: Mass spectrum $dN/(dy_d dM^2)$ integrated over the transverse 
momentum $p_T$ of the pair for $p_T > 8$ GeV/$c$.
{\it Right}: Transverse momentum spectrum $dN/(dy_d d^2 p_T)$ integrated over 
a mass range 0.5 GeV $< M<$ 1 GeV. 
Both panels show the case $y_d=0$ for the pair rapidity $y_d$ and a cut 
$|y_e| < 0.5$ for the single electron rapidity.}
\end{center}
\end{figure}

This work was supported in parts by DOE grants DE-FG02-87ER40328,
DE-AC02-98CH10886, RIKEN/BNL, the Texas A\&M College of Science,
and the Natural Sciences and Engineering Research Council of Canada.\\[0.5em]



\begin{thebibliography}{99}

\bibitem{Fries:2002kt}
  Fries R J, M\"uller B and Srivastava D K 2003
  {\it Phys.\ Rev.\ Lett.}  {\bf 90} 132301

\bibitem{Zakharov:2004bi}
  Zakharov B G 2004
  {ETP Lett.}  {\bf 80}  1

\bibitem{Turbide:2005fk}
  Turbide S, Gale C, Jeon S and Moore G D 2005
  {\it Phys.\ Rev.} C {\bf 72} 014906

\bibitem{Turbide:2006mc}
  Turbide S, Gale C, Srivastava D K and Fries R J 2006
  {\it Phys.\ Rev.}  C {\bf 74} 014903




\end{thebibliography}
\end{document}